\documentclass[12pt]{article}
\usepackage[dvips]{graphicx}
\usepackage{amsmath,amssymb}
\usepackage{color}

\baselineskip=\normalbaselineskip

\newcommand{\be}{\begin{equation}}
\newcommand{\ee}{\end{equation}}
\newcommand{\bea}{\begin{eqnarray}}
\newcommand{\eea}{\end{eqnarray}}
\newcommand{\ba}{\begin{array}}
\newcommand{\ea}{\end{array}}

\newcommand{\tr}{\mbox{tr}}

\newcommand{\cG}{{\cal{G}}}

\newcommand{\cM}{{\cal{M}}}

\newcommand{\cO}{{\cal{O}}}

\newcommand{\nn}{\nonumber \\}

\newcommand{\eq}[1]{(\ref{#1})}

\newcommand{\tn}{\widetilde{n}}

\begin{document}

\begin{titlepage}
\begin{flushright}
\normalsize {\tt hep-th/0404204}\\  
April 27, 2004 \\
RIKEN-TH-19
\end{flushright}

\vfill

\begin{center}
{\large \bf
String Theoretical Interpretation 
for \\
Finite $N$ Yang-Mills Theory in Two-Dimensions
}

\vfill

{Toshihiro Matsuo\footnote{\tt tmatsuo@riken.jp} and
So Matsuura\footnote{\tt matsuso@riken.jp}}

\vfill
{\it
Theoretical Physics Laboratory,\\
The Institute of Physical and Chemical Research (RIKEN),\\
Wako, Saitama, 351-0198, Japan
}

\end{center}

\vfill
\begin{abstract}
We discuss the equivalence between a string theory and
the two-dimensional Yang-Mills theory 
with $SU(N)$ gauge group for finite $N$. 
We find a sector which can be interpreted as 
a sum of covering maps from closed string world-sheets 
to the target space, 
whose covering number is less than $N$.
This gives an asymptotic expansion of $1/N$  
whose large $N$ limit becomes the chiral sector defined 
by D.~Gross and W.~Taylor. 
We also discuss that the residual part of the partition function
provides the non-perturbative corrections to the perturbative
expansion.

\end{abstract}
\vfill

\end{titlepage}

\setcounter{footnote}{0}

\section{Introduction}

Recent studies on string theories have shown 
that exploring the relationship among string theories, 
gauge theories and matrix models 
is an important issue for understanding the string theories 
themselves. 
As typical examples, the duality between gauge theories and 
string theories \cite{tHooft} is realized as 
the AdS/CFT correspondence \cite{ads-cft} 
(for a recent review, see also Ref.\,\cite{FMS}), 
while the large $N$ reduction of gauge theories realizes 
the gauge/matrix correspondence \cite{EK},
and large $N$ matrix models have been proposed as candidates 
for non-perturbative definition of the string/M theory
\cite{Ishibashi:1996xs, Banks:1996vh}. 

Explicit analysis is rather tractable in lower dimensions.
The duality between two-dimensional non-critical string theory and 
the $c=1$ matrix model is one of the most successful examples.
It has recently been found that the non-perturbative effect 
predicted from the $c=1$ matrix model indeed 
describes D-branes 
in the non-critical string theory \cite{c=1}.%
\footnote{ 
For other discussions for the matrix/string correspondence, 
see, e.g. Refs.\,\cite{Chaudhuri:1993zr,Nakajima:1998vj}.} 
As for the gauge/string equivalence, Gross and Taylor
have attempted in their seminal works \cite{G,GT,GT2} to uncover 
the relationship between 
the two-dimensional Yang-Mills theory (YM$_2$) and a string theory, 
to which we pay particular attention in this paper. 
The statement of this equivalence is that 
the free energy of Yang-Mills theory 
is identified with a partition function of a string theory 
on the target space $\cM$ with the string coupling $1/N$ 
and the string tension $\lambda$ \cite{G}.
Strong evidence for this conjecture 
is given in Refs.\,\cite{G}--\cite{GT2}  
by showing that the partition function of the large $N$
YM$_2$ can be interpreted as a sum of all maps without folds 
from a set of closed string world-sheets 
to the target space. 
(For review and further developments, see,
e.g. Refs.\,\cite{CMR}--\cite{CDMP}.)
This means that YM$_2$ describes at least 
a perturbative closed string theory in the sense that 
the asymptotic $1/N$ expansion of the free energy
corresponds to a perturbative expansion of a closed string theory.

In this paper, we are interested in a string theoretical
interpretation of the finite $N$ YM$_2$. 
The crucial point is that we know the exact expression of the
partition function of the YM$_2$ for finite $N$.
According to the modern understanding of critical and/or 
non-critical string theories, a closed string theory contains 
not only closed strings themselves but also open string degrees 
of freedom emerging through D-branes, 
which are the non-perturbative objects
characterized by a factor $e^{-1/g_s}$
in free energy of the theory. 
It is natural to wonder whether this is the case or not  
for a string theory defined through YM$_2$. 
We show that the partition function of YM$_2$ takes
a form which enable us to conclude that this is indeed the case  
by dividing the partition function into two sectors in a definite way. 
One of these sectors turn out to be interpreted as 
a sum of maps from closed string world-sheets, 
whose large $N$ limit 
becomes the chiral sector \cite{GT} of the large $N$ theory. 
Furthermore, 
we argue that the other sector has properties that 
is consistent with non-perturbative effects of a closed string theory.

This paper is organized as follows. 
In the next section we briefly review the relationship between 
YM$_{2}$ and a string theory. 
We describe here the basic ingredients and concepts 
needed to understand the relationship.
In section 3, 
we examine the partition function of the finite $N$ YM$_2$ 
and discuss the string theoretical interpretation from the perturbative 
and the non-perturbative points of view.  
Section 4 is devoted to summarize our results and discussions.
In the appendix, we review a geometric interpretation of 
the large $N$ YM$_{2}$ partition function.

\section{YM$_2$/string correspondence}
\setcounter{equation}{0}

In this section, we briefly review the duality between YM$_2$
and a string theory 
\cite{G}--\cite{GT2}. 
First, we will explain the technique for translating the partition function 
of YM$_2$ into an expression 
of which we can give a geometrical interpretation. 
We will next show that the resulting expression can further be 
regarded as a perturbative expansion of a closed string theory, 
especially in the case of large $N$ theory.

Let us consider the Yang-Mills theory with $SU(N)$ gauge symmetry 
on a two-dimensional (Euclidean) manifold $\cM$ 
with genus $G$. 
This theory is exactly solvable 
and the partition function is evaluated as \cite{Migdal} 
\bea
Z_{\rm YM}(G, \lambda A, N)
&=&\int [dA_{\mu}] \exp\left[
{{-\frac{1}{4g_{YM}^2}\tr\int d^2x
\sqrt{g}F_{\mu\nu}F^{\mu\nu}}}\right] \nn 
&=&\sum_R (\dim R)^{2-2G} \exp\left[-\frac{\lambda A}{2N} C_2(R)\right], 
\label{YM2}
\eea
where 
$\lambda\equiv g_{\rm YM}^2 N$ is the 't Hooft coupling, 
$A$ is the area of the manifold $\cM$, 
$R$ is an irreducible representation of the gauge group, 
and $\dim R$ and $C_2(R)$ are the dimension and the quadratic Casimir of $R$, respectively.

In the following, we will rewrite the partition function \eq{YM2} 
in the language of the symmetry group. 
The irreducible representation of $SU(N)$ 
can be classified by the Young diagrams with the rows less than $N$, 
and that of the symmetry group $S_n$ can be classified 
by the Young diagrams with $n$ boxes. 
Thus, we will use the same notation $R$ to 
express the Young diagram corresponding to the 
irreducible representation of $SU(N)$ and $S_n$.

One might conclude that the partition function 
in the large $N$ theory is given by 
\begin{align}
\sum_{n=1}^{\infty}\sum_{R\in Y_n}
(\dim R)^{2-2G} \exp\left[-\frac{\lambda A}{2N} C_2(R)\right], 
\end{align}
where $Y_n$ denotes the set of Young diagrams with $n$ boxes. 
However, as discussed in Ref.\,\cite{GT}, this is a half of the
full large $N$ theory, 
since the contribution to the partition function 
of a Young diagram $R$ with a finite number of boxes 
is the same as that of the conjugate representation $\bar{R}$.  
To take the both contributions into account, the authors of Ref.\,\cite{GT} 
conjecture that the $1/N$ expansion of 
the partition function is evaluated from  
\begin{align}
 Z_{\rm YM}(G,\lambda A,N)=
 \sum_{n=1}^{\infty} \sum_{\tn=1}^{\infty}
 \sum_{R\in Y_n} \sum_{S\in Y_{\tn}}(\dim \overline{S}R)^{2-2G}
 e^{-\frac{\lambda A}{2N} C_2(\overline{S}R)}, 
 \label{large-N}
\end{align}
where $\bar{S}R$ is called the composite diagram made from 
Young diagrams $R$ and $S$, 
whose column length is 
\begin{equation}
\begin{cases}
N-\tilde{c}_{L+1-i}, \qquad &i \leq L ,
\nn
c_{i-L}, \qquad &i>L ,
\end{cases}
\end{equation}
where $c_{i}$ and $\tilde{c}_{i}$ are the length of $i$-th column
of $R$ and $S$, respectively, and $L$ is the length of the first row 
of the representation $S$.
Since the dimension and the quadratic Casimir of the composite 
representation are 
\begin{align}
 \dim{\overline{S}R} &= \dim{R}\dim{S}
  \left[1+\cO\left(1/N^2\right)\right], \\
 C_2(\overline{S}R) &= C_2(R) + C_2(S) + \frac{2n\tn}{N}, 
\end{align}
we can see that the partition function of the full theory 
is factorized into two copies of a chiral sectors \cite{GT}, 
\begin{align}
 Z^+(G,\lambda A,N)\equiv 
 \sum_{n=1}^{\infty}\sum_{R\in Y_n}
  (\dim R)^{2-2G} \exp\left[-\frac{\lambda A}{2N} C_2(R)\right], 
 \label{chiral}
\end{align}
except for the $\cO(1/N^2)$ correction of the dimension 
and a coupling term $e^{-\frac{\lambda A n\tn}{N^2}}$ from 
the quadratic Casimir. 
In the following analysis, we concentrate the discussion only on the chiral
sector. For a geometrical interpretation of the full partition
function \eq{large-N}, see Ref.\,\cite{GT2}.

The dimension and the quadratic Casimir 
of the irreducible representation $R$ are evaluated by   
using the character of the symmetric group $S_n$ \cite{GT2};
\begin{align}
 \label{dim}
 \dim R &= \frac{1}{n!}
 \sum_{\sigma\in S_n}{\chi_R(\sigma)}{N^{K_{\sigma}}} 
 \equiv \frac{d_R N^n}{n!}\frac{\chi_R(\Omega_n)}{d_R}, \\
 C_2(R) &= Nn + 2\sum_{p\in T_2}\frac{\chi_R(p)}{d_R} - \frac{n^2}{N}\,,
 \label{Casimir}
\end{align}
where 
$\chi_R(\sigma)$ is the character of $\sigma\in S_n$ 
in the representation $R$, 
$d_R$ is the dimension of the representation $R$ of $S_n$, 
$T_2$ is the set of transpositions, 
and 
$\Omega_n \equiv \sum_{\sigma\in S_n}
{\sigma}/{N^{n-K_\sigma}}$ where $K_\sigma$ 
is the number of cycles in $\sigma$, which
is an element of the group ring of $S_n$. 
Substituting these into the partition function \eq{chiral},
one will find that the partition function of the chiral sector becomes%
\footnote{
We have used the formulae
\begin{align}
 \sum_{\sigma \in T}\chi_R(\sigma)\chi_R(\rho)
   &=d_R \sum_{\sigma \in T} \chi_R(\sigma\rho) 
 \quad (T:\text{a conjugacy class in $S_n$}), \nn 
 \sum_{s,t\in S_n}\frac{\chi_R(sts^{-1}t^{-1})}{d_R}
   &=\biggl(\frac{n!}{d_R}\biggr)^2, \quad 
   \mbox{and} \quad
  \delta(\sigma) = \frac{1}{n!} \sum_{R\in Y_n}d_R \chi_R(\sigma)\,. 
 \nonumber
\end{align}
The definition of the delta function in $S_n$ is 
\[
 \delta(\sigma) = 
 \begin{cases}
  1 & \sigma = 1 \nn
  0 & {\rm others}
  \end{cases}
  \quad {\rm for}\,\, \sigma\in S_n. 
\]
}
\begin{align}
 Z^+(G,\lambda A,N)
 &= \sum_{n=1}^{\infty}e^{-\frac{n\lambda A }{2}}
 \sum_{i,t,h=0}^{\infty}
  \frac{(-1)^i\left(\lambda A\right)^{i+t+h}}{i!\,t!\,h!}
  \left(\frac{n(n-1)}{2}\right)^t\left(\frac{n}{2}\right)^h
  N^{n(2-2G)-i-2t-2h} \nn
 &\times\sum_{s_1,t_1,\cdots,s_G,t_G \in S_n}\sum_{p_1,\cdots,p_i\in T_2}
 \frac{1}{n!}\delta(
  p_1\cdots p_i \Omega_n^{2-2G} \prod_{j=1}^G 
  s_j t_j s_j^{-1} t_j^{-1} )\,. 
 \label{chiral2}
\end{align}
This is the expression of the partition function 
in which a geometrical interpretation is possible. 
In fact, \eq{chiral2} can be interpreted as a 
sum of covering maps from a set of two dimensional manifolds   
to the target space. 
The indices $i$, $t$ and $h$ is regarded as the number of 
single branch points, tubes, and contracted handles \cite{GT}, respectively.
Furthermore the delta function in \eq{chiral2} provides 
a sum of the symmetry factor 
$|S_{\nu}|$ of a given covering map $\nu$ in a set of covering maps $\Sigma(G,n,i)$;
\bea
\sum_{s_1,t_1,\cdots,s_G,t_G \in S_n}\sum_{p_1,\cdots,p_i\in T_2}
 \frac{1}{n!}\delta(
  p_1\cdots p_i \Omega_n^{2-2G} \prod_{j=1}^G 
  s_j t_j s_j^{-1} t_j^{-1} )
  =\sum_{\nu \in \Sigma(G,n,i)}1/|S_{\nu}| .
\eea
Details of the geometrical interpretation are provided 
in the appendix.

In order to obtain the free energy $W^+(G,\lambda A, N)$ 
which is the logarithm of 
the partition function, it is enough to take 
restricted sums in the partition function which contain a set of 
connected part of the covering maps $\tilde{\Sigma}(G,n,i)$ ;
\begin{align}
 W^+(G,\lambda A,N)
 &= \sum_{n=1}^{\infty}e^{-\frac{n\lambda A }{2}}
 \sum_{i,t,h=0}^{\infty}
  \frac{(-1)^i\left(\lambda A\right)^{i+t+h}}{i!\,t!\,h!}
  \left(\frac{n(n-1)}{2}\right)^t\left(\frac{n}{2}\right)^h
   \nn
 &\times N^{n(2-2G)-i-2t-2h}
 \sum_{\nu \in \tilde{\Sigma}(G,n,i)}1/|S_{\nu}|. 
 \label{chiral free energy}
\end{align}

To complete our discussion, we must further show that 
the above geometrical interpretation is also 
``string theoretical'', that is, 
the expression \eq{chiral free energy} 
can be interpreted as a sum of maps 
from closed string world-sheets to the target space with a weight 
corresponding to a world-sheet action. 
First of all, the term $e^{-\frac{n\lambda A}{2}}$  
is naturally interpreted as a contribution from a Nambu-Goto type 
world-sheet action if the 't Hooft coupling $\lambda$ is identified with
the string tension. 
Next, we would like to interpret $1/N$ as a string coupling $g_s$, 
however, there is a subtlety here.
Since the Euler characteristic of a closed string world-sheet is even number,
we must show that \eq{chiral} contains only even powers of $1/N$.
We will see below that \eq{chiral free energy} 
is indeed the case 
although it seems to contain odd powers of $1/N$.

As mentioned in Ref.\,\cite{G}, if $N$ is large enough 
there exists the ``transpose representation'' $R^T$
for any irreducible representation $R$ in the sum in \eq{YM2}.  
Here, the Young diagram $R^T$ is defined by exchanging 
the rows and columns of the original diagram $R$.  
When the dimension and the quadratic Casimir of the representation $R$ 
are given as \eq{dim} and \eq{Casimir}, respectively, 
those of $R^T$ are \cite{G} 
\begin{align}
 \dim R^T &= \frac{N^n}{n!}
 \sum_{\sigma\in S_n} (-N)^{K_{\sigma}-n}\chi_R(\sigma), \\
 C_2(R^T) &= Nn - 2\sum_{p\in T_2}\frac{\chi_R(p)}{d_R} - \frac{n^2}{N}\,.
\end{align}
Thus, by rewriting the partition function \eq{chiral} as 
\begin{align}
Z^+&(G, \lambda A, N)  \nn
 &=\sum_{n=1}^{\infty}
  \sum_{R\in Y_n}\frac{1}{2}\biggl\{
  (\dim R)^{2-2G} 
  \exp\left[-\frac{\lambda A}{2N} C_2(R)\right] 
  +(\dim R^T)^{2-2G} 
  \exp\left[-\frac{\lambda A}{2N} C_2(R^T)\right]
  \biggr\}\,, 
 \label{even}
\end{align} 
we can see that only even powers of $1/N$ survive. 
For example, let us write down the partition function for $G=1$ 
in the language of $S_n$ as follows:  
\begin{align} 
Z^+&(G=1, \lambda A, N) \nn
 &= \sum_{n=1}^{\infty} e^{-\frac{n\lambda A}{2}}
  \sum_{c,t,h=0}^{\infty}\frac{(\lambda A)^{2c+t+h}}{(2c)!\,t!\,h!}
  \left(\frac{n(n-1)}{2}\right)^t\left(\frac{n}{2}\right)^h
  N^{-2(c+t+h)}  \nn 
  &\qquad \times\sum_{s,t \in S_n}\sum_{p_1,\cdots,p_{2c}\in T_2}
  \frac{1}{n!}\delta(
  p_1\cdots p_{2c}
  s t s^{-1} t^{-1} )\,.
 \label{chiral3}
\end{align} 
As announced, the exponents of $N$ contain only even numbers. 
In \eq{chiral3}, the index $c$ can be regarded as the number 
of cuts on the world-sheets. 
It is straightforward to carry out the same calculation 
for an arbitrary number of $G$.

\section{
Dual string theory for the finite $N$ YM$_2$
}
\setcounter{equation}{0}

In the previous section, we have reviewed the geometrical and 
string theoretical interpretation of YM$_2$ given by Gross and Taylor.
From their analysis, it seems to be natural to conclude that 
YM$_2$ describes a closed string theory. 
However, there are two subtle points in the analysis 
if we consider the non-perturbative corrections; 
first, the way to decompose a Young diagram $T$ into 
$S$ and $R$ as $T=\overline{S}R$ is not unique,
and second, in \eq{chiral}, 
there are diagrams in $Y_n$ that do not correspond 
to any irreducible representation of $SU(N)$ 
when $n$ is larger than $N$. 
Although we do not face
these problems in large $N$ limit, 
we must start with finite $N$ YM$_2$ to evaluate 
the $1/N$ perturbative expansion and the non-perturbative corrections 
of the partition function \eq{YM2}. 
Actually, a geometrical interpretation of the 
finite $N$ YM$_2$ is also possible \cite{BT}.  
However, as we will mention later, 
this interpretation is not satisfactory to claim 
that the finite $N$ YM$_2$ is equivalent 
to a string theory. 
In this section, we will refine the discussion in Ref.\,\cite{BT} 
and give a string theoretical interpretation of the partition function 
for the finite $N$ case.   
We will also mention 
the connection between the finite $N$ theory and the large $N$ theory.

Let us first recall the geometrical interpretation for the finite $N$ YM$_2$ 
given in Ref.\,\cite{BT}. 
Using the same technique as that for the large $N$ theory, 
we can rewrite the partition function of the finite $N$ theory \eq{YM2}  
as
\begin{align}
 Z_{\rm YM}(G,\lambda A,N)
 &= \sum_{n=1}^{\infty}e^{-\frac{n\lambda A }{2}}
 \sum_{i,t,h=0}^{\infty}
  \frac{(-1)^i\left(\lambda A\right)^{i+t+h}}{i!\,t!\,h!}
  \left(\frac{n(n-1)}{2}\right)^t\left(\frac{n}{2}\right)^h
  N^{n(2-2G)-i-2t-2h} \nn
 &\times\sum_{s_1,t_1,\cdots,s_G,t_G \in S_n}\sum_{p_1,\cdots,p_i\in
 T_2}
 \sum_{R\in Y_n^{N}}
 \frac{d_R}{(n!)^2}\chi_R(
  p_1\cdots p_i \Omega_n^{2-2G} \prod_{j=1}^G 
  s_j t_j s_j^{-1} t_j^{-1} )\,, 
 \label{finite-N}
\end{align}
where $Y_n^N$ is the set of Young diagrams with $n$ boxes 
and less than $N$ rows. 
Note that $Y_n^N$ is equal to $Y_n$ if $n$ is less than $N$. 
To interpret (\ref{finite-N}) geometrically, we must express the sum of 
representations in a form of the delta function of products 
of elements of $S_n$. 
To this end,
the authors of Ref.\,\cite{BT} defined 
the projection operator:%
\footnote{
Use of the formula
\begin{equation}
 \sum_{\sigma\in S_n}\chi_R(\sigma)\chi_R(\sigma^{-1}\tau)
  =\frac{n!}{d_R}\chi_R(\tau), \nonumber 
\end{equation}
makes it straightforward to show that $P_n^{(N)}$ is indeed 
a projection operator, that is, $\left(P_n^{(N)}\right)^2=P_n^{(N)}$.
} 
\begin{equation}
 P_n^{(N)}\equiv \sum_{\rho\in S_n}\sum_{R\in Y_n^N}\frac{d_R}{n!}
  \chi_R(\rho)\rho
  \equiv \sum_{\rho\in S_n}N^{K_\rho-n}P^{(N)}(\rho)\rho,
\end{equation}
which has the property
\begin{equation}
 \delta(\sigma P_n^{(N)})
  =\sum_{R\in Y_n^N}\frac{d_R}{n!}\chi_R(\sigma), 
\end{equation}
for any $\sigma\in S_n$.
Then \eq{finite-N} can be rewritten as 
\begin{align}
 Z_{\rm YM}(G,\lambda A,N)
 &= \sum_{n=1}^{\infty}e^{-\frac{n\lambda A }{2}}
 \sum_{i,t,h=0}^{\infty}
  \frac{(-1)^i\left(\lambda A\right)^{i+t+h}}{i!\,t!\,h!}
  \left(\frac{n(n-1)}{2}\right)^t\left(\frac{n}{2}\right)^h
  N^{n(2-2G)-i-2t-2h} \nn
 &\times\sum_{s_1,t_1,\cdots,s_G,t_G \in S_n}\sum_{p_1,\cdots,p_i\in
 T_2}
  \frac{1}{n!}\delta(
  p_1\cdots p_i \Omega_n^{2-2G} P_n^{(N)} \prod_{j=1}^G 
  s_j t_j s_j^{-1} t_j^{-1} )\,. 
 \label{finite-N2}
\end{align}
The only difference between this expression and the chiral sector of 
the large $N$ theory \eq{chiral2} 
is the existence of the projection operator 
$P_n^{(N)}$ in the delta function. 
In Ref.\,\cite{BT}, the projection operator is interpreted 
as a sum of multiple branch points (the projection point) 
on the covering sheets, 
and thus, an element $\rho$ in $P_n^{(N)}$ contributes 
$(K_\rho-n)$ to the Euler characteristic.%
\footnote{
In Ref.~\cite{BT}, it is discussed that 
the $N^{n-K_\rho}$ which compensates the $N$ dependence 
comes from an unknown string interaction. 
} 
Therefore, it is possible to give a geometrical interpretation 
for the partition function \eq{finite-N2}.

However, 
there are some difficulties in claiming 
that finite $N$ YM$_2$ is equivalent to a closed string theory 
through the above interpretation:
first, 
the partition function \eq{finite-N} contains the terms 
with odd powers of $1/N$;
second, we do not have any candidate for a string interaction 
which provides a contribution $N^{n-K_\rho}$ to the partition function;  
and third, the relationship to the large $N$ theory is not clear. 
We here give a way to overcome these difficulties 
by separating the partition function of the finite $N$ YM$_2$ 
into two sectors, which we call ``perturbative sector'' and 
``residual sector'', as
\begin{align}
 Z_{\rm YM}(G,\lambda A,N) = Z_{\rm YM}^{\rm pert}(G,\lambda A,N) 
                            +Z_{\rm YM}^{\rm res}(G,\lambda A,N), 
\end{align}
where 
\begin{align}
 \label{cutoff}
 Z_{\rm YM}^{\rm pert}(G,\lambda A,N) &\equiv 
 \sum_{n=1}^{N-1}\sum_{R\in Y_n}
 \left(\dim R\right)^{2-2G}
 \exp\left[-\frac{\lambda A}{2N}C_2(R)\right], \\
 Z_{\rm YM}^{\rm res}(G,\lambda A,N) &\equiv 
 \sum_{n=N}^{\infty}\sum_{R\in Y_n^N}
 \left(\dim R\right)^{2-2G}
 \exp\left[-\frac{\lambda A}{2N}C_2(R)\right]. 
 \label{residual}
\end{align} 
For the purposes of subsequent discussion, 
we will translate \eq{cutoff} and \eq{residual} 
into the language of the symmetry group for $G=1$:
\begin{align}
  \label{cutoff2}
  Z_{\rm YM}^{\rm pert}(G=1,\lambda A,N)
 &= \sum_{n=1}^{N-1} e^{-\frac{n\lambda A}{2}}
  \sum_{c,t,h=0}^{\infty}\frac{(\lambda A)^{2c+t+h}}{(2c)!\,t!\,h!}
  \left(\frac{n(n-1)}{2}\right)^t\left(\frac{n}{2}\right)^h
  N^{-2(c+t+h)}  \nn 
  &\qquad \times\sum_{s,t \in S_n}\sum_{p_1,\cdots,p_{2c}\in T_2}
  \frac{1}{n!}\delta(
  p_1\cdots p_{2c}
  s t s^{-1} t^{-1} )\, \\
 Z_{\rm YM}^{\rm res}(G=1,\lambda A,N)
 &= \sum_{n=N}^{\infty}e^{-\frac{n\lambda A }{2}}
 \sum_{i,t,h=0}^{\infty}
  \frac{(-1)^i\left(\lambda A\right)^{i+t+h}}{i!\,t!\,h!}
  \left(\frac{n(n-1)}{2}\right)^t\left(\frac{n}{2}\right)^h
  N^{-i-2t-2h} \nn
 &\times\sum_{s,t \in S_n}\sum_{p_1,\cdots,p_i\in
 T_2}
 \sum_{R\in Y_n^{N}}
 \frac{d_R}{(n!)^2}\chi_R(
  p_1\cdots p_i  s t s^{-1} t^{-1} )\,. 
 \label{residual2}
\end{align}
To derive \eq{cutoff2}, we have used the fact that there exists 
the transpose representation $R^T$ 
for any representation $R$ in \eq{cutoff}, as in the case of the chiral
sector \eq{chiral}.

The partition function \eq{cutoff} contains only even powers of $1/N$, 
thus we can interpret this as a sum over covering maps from a set of
closed string world-sheets  
by repeating the consideration described in the previous section. 
This sector takes the same form as 
that of the chiral sector of the large $N$ theory,
except that the range for the summation is now restricted to $N\!-\!1$.
This means that \eq{cutoff} corresponds to the perturbative 
expansion of a closed string theory, which is the reason 
why we refer to \eq{cutoff} as the ``perturbative'' sector. 
The restriction of the range for the summation tells us that 
there is an upper bound to the wrapping number of the world-sheets, 
which is a manifestation of the stringy exclusion principle 
\cite{Maldacena:1998bw}. 
Moreover, the coefficients of the $1/N$ expansion defined by
\eq{cutoff} become those of the chiral sector \eq{chiral2} 
in the large $N$ limit. 
From these facts, we claim that the perturbative sector of the partition
function in the finite $N$ case 
corresponds to the chiral sector of the full
large $N$ theory. 

On the other hand, we cannot give a satisfactory 
closed string interpretation of the residual sector 
since odd powers of $1/N$ exist in this sector. 
However, it is natural to consider that 
the residual sector 
should contain corrections to the perturbative 
expansion described by the perturbative sector. 
In fact, there are some signs in the explicit expression \eq{residual2} 
which support this expectation. 
The most suggestive is the overall factor 
$e^{-\frac{N\lambda A}{2}}$ in the free energy%
\footnote{
It is easy to see that the overall factor $e^{-N\lambda A/2}$ 
is ubiquitous in the connected part of the residual sector
partition function.
}. 
This is compatible with the fact that 
non-perturbative effects of a string theory appear 
to the string partition function as $e^{-\cO(1/g_s)}$. 
Moreover, recent developments in the non-perturbative aspects of
string theories have elucidated that the 
non-perturbative effects come from D-branes. 
Since there are open strings on the D-branes, 
there should be odd powers of $g_s$ in the partition function 
of string theory if we take into account the non-perturbative effects. 
From this observation, 
the presence of the odd powers of $1/N$ in \eq{residual} would 
rather suggest that we should consider open string world-sheets 
to give a string theoretical interpretation of the residual sector. 
The above considerations lead us to conjecture that 
the residual sector \eq{residual2} contains contributions 
from open string world-sheets.%
\footnote{
For another discussion on the non-perturbative string effects 
in YM$_2$, see Ref.\,\cite{LMR}. 
} 
To prove this, however, 
we must show that the residual sector 
indeed contains information which can be interpreted as 
a map from open string world-sheets.
This is important future work \cite{MM2}.

We conclude this section by considering another way 
to divide the partition function \eq{finite-N}. 
If one respects the even powers of $1/N$ in the partition function, 
one may restrict the sum of representations to those whose corresponding
Young diagrams can be put in the $N\times N$ squared box. 
Let us express the set of such diagrams with $n$ boxes as 
$Y_n^{N\times N}$.
Since any Young diagram $R$ in $Y_n^{N\times N}$ has 
the transpose diagram $R^T$ in $Y_n^{N \times N}$, 
the restricted partition function becomes ($G=1$, for simplicity)
\begin{align}
Z_{\rm YM}^{N\times N}(G=1, \lambda A, N)  
  &= \sum_{n=1}^{N^2}
  \sum_{R\in Y_n^{N\times N}}
  \exp\left[-\frac{\lambda A}{2N} C_2(R)\right]   \nn 
  &= \sum_{n=1}^{N^2} e^{-\frac{n\lambda A}{2}}
  \sum_{c,t,h=0}^{\infty}\frac{(\lambda A)^{2c+t+h}}{(2c)!\,t!\,h!}
  \left(\frac{n(n-1)}{2}\right)^t\left(\frac{n}{2}\right)^h
  N^{-2(c+t+h)}  \nn 
  &\qquad \times\sum_{s,t \in S_n}\sum_{p_1,\cdots,p_{2c}\in T_2}
  \sum_{R\in Y_n^{N\times N}}\frac{d_R}{(n!)^2}\chi_R(
  p_1\cdots p_{2c}
  s t s^{-1} t^{-1} )\,. 
 \label{NxN}
\end{align}
However, to provide a geometrical interpretation, 
we must introduce a projection operator, 
\begin{equation}
  P_n^{N\times N}\equiv \sum_{\rho\in S_n}
   \sum_{R\in Y_n^{N\times N}}\frac{d_R}{n!}
   \chi_R(\rho)\rho, 
   \label{projection-NxN}
\end{equation}
and rewrite \eq{NxN} as 
\begin{align}
  Z_{\rm YM}^{N\times N}(G=1,\lambda A,N)
 &= \sum_{n=1}^{N^2} e^{-\frac{n\lambda A}{2}}
  \sum_{c,t,h=0}^{\infty}\frac{(\lambda A)^{2c+t+h}}{(2c)!\,t!\,h!}
  \left(\frac{n(n-1)}{2}\right)^t\left(\frac{n}{2}\right)^h
  N^{-2(c+t+h)}  \nn 
  &\qquad \times\sum_{s,t \in S_n}\sum_{p_1,\cdots,p_{2c}\in T_2}
  \frac{1}{n!}\delta(
  P_n^{N\times N}p_1\cdots p_{2c}
  s t s^{-1} t^{-1} )\,. 
\end{align}
Since, we interpret an element of $S_n$ in 
the delta function as an existence of a multiple branch point 
on the covers, it is natural to assume that the factor 
$N^{-n+K_\rho}$ accompanies the element $\rho\in S_n$ in 
the projection operator \eq{projection-NxN} 
like as in the case of the finite $N$ partition function \eq{finite-N2}.
Thus, terms with odd powers of $1/N$ still remain 
in the restricted partition function \eq{NxN}, which
is hard to be interpreted as a perturbative closed string theory. 
This analysis indicates that the way of dividing the partition function 
that we have adopted in this paper would be a more proper way 
to separate the perturbative and non-perturbative effects of 
the dual string theory.

\section{Conclusion and discussion}
In this paper, we discussed the equivalence of the finite $N$ YM$_2$ 
to a string theory. 
YM$_2$ with the gauge group $SU(N)$ 
is exactly solvable and the partition function is expressed 
as a sum of a function of the irreducible representation $R$ of $SU(N)$ 
(see \eq{YM2}).
We separated the partition function into two sectors by introducing a cutoff
parameter $N$ to the number of boxes of the corresponding Young diagram. 
We demonstrated that the sector that is defined by the
sum of diagrams with boxes less than $N$ (the perturbative sector) 
corresponds to the perturbative expansion of a closed string theory. 
The wrapping number of the closed strings that constitute this sector 
is restricted to less than $N$, 
which is a manifestation of the stringy exclusion principle. 
We also showed that this sector becomes the chiral sector \cite{GT} 
in the large $N$ limit. 
We presented the conjecture that we can also give a string theoretical 
interpretation of the other sector (the residual sector) 
by taking into account non-perturbative effects of the dual string theory. 
We proposed some evidence in support of this conjecture.

Finally, we make some comments. 
Although we defined the finite $N$ counterpart of 
one of the chiral sectors of the large $N$ theory, 
we do not know yet how the other chiral sector 
appears as a large $N$ limit of the finite $N$ YM$_2$. 
Recalling the way of construction of the composite diagram, 
we see that one of the chiral sectors comes from a sum over  
the set of Young diagrams $\{R\}$ with a finite number of boxes and  
the other chiral sector comes from a sum over the set of 
the conjugate diagrams $\{\bar{S}\}$ of Young diagrams $\{S\}$ with 
a finite number of boxes. 
In the large $N$ theory, we must take these summation 
independently since $\bar{S}$ has an infinite number of boxes. 
For the finite $N$ theory, however, the number of boxes in $\bar{S}$ 
is also finite, and thus, the summation in \eq{YM2} would contain 
both of the chiral sectors. 
From this consideration, 
it is plausible to assume that the residual sector \eq{residual} 
also contains the finite $N$ counterpart of the other chiral sector,  
not only non-perturbative corrections to the perturbative expansion. 
In order to show this, we must separate contributions from perturbative 
and non-perturbative effects of the dual string theory 
in the residual sector. 
As one possibility, non-perturbative effects could be 
evaluated as a sum over open string world-sheets whose boundaries 
shrink to points.  
This is expected because D-branes would excite on the target space 
if the string coupling constant is finite 
and they make holes to the world-sheets. 
Note that the boundaries must shrink to points  
because of the area preserving diffeomorphism invariance of YM$_2$. 
Another advantage of assuming the D-branes is that 
they might suppress the wrapping number of world-sheets 
because the appearance of D-branes would decrease 
the free energy of the system 
when the wrapping number of the world-sheets becomes too large. 
This might provide a reason for the string exclusion principle 
being realized in the dual string theory of the finite $N$ YM$_2$. 
We hope to confirm this scenario and 
to report subsequent developments in the near future \cite{MM2}.

\section*{Acknowledgement}
The authors would like to thank H.~Kawai, T.~Tada, M.~Hayakawa, 
T.~Kuroki, S.~Kawamoto, K.~Ohta, Y.~Shibusa and Y.~Hikida 
for useful discussions and valuable comments. 
This work has been supported by 
the Special Postdoctoral Researchers Program at RIKEN.

\appendix
\section{Geometric interpretation of the YM$_2$ partition function}
In this appendix, we review the geometric interpretation 
for the partition function of the chiral sector. 
For a complete discussion, see Refs.\,\cite{GT,GT2}. 

The starting point is the partition function \eq{chiral2}. 
In the following, show that \eq{chiral2} is equal 
to a sum over covering maps from a set of (disconnected) 
two-dimensional manifolds 
to the target space $\cM$ \cite{G,GT,GT2}. 
As a preparation, let us consider a n-fold covering map $\nu$ 
with $i$ single branch points at $\{w_1,\cdots,w_i\}\in \cM$ 
and $|2-2G|$ multiple branch points (twists points) at 
$\{z_1,\cdots,z_{|2-2G|}\}\in \cM$. 
We assume that $\nu$ also maps some handles 
of the sheets to points on $\cM$.%
\footnote{  
This map must not break 
the area preserving diffeomorphism of 
YM$_2$. } 
Since there are $n$ sheets over a point on the target space, 
we classify such a mapped handle into two types; 
one connects two different sheets (a tube), 
and the other is on a same sheet (a contracted handle) \cite{GT}. 
Here, we fix the number of tubes and contracted handles to 
$t$ and $h$, respectively. 
In addition, we assume that the positions of the branch points, 
the tubes and the contracted handles can move over $\cM$,
while those of the multiple branch points are fixed.

Under the above setting, we evaluate 
(1) the sum of the Euler characteristic of the sheets on $\cM$, 
(2) the numerical factor that comes from integrating over 
the parameters of the map $\nu$, 
and 
(3) the symmetry factor of the map. 

\begin{description}
\item (1) Euler characteristic

If there are no branch points, tubes and contracted handles, 
the Euler characteristic of the sheet is $n(2-2G)$ 
since the map $\nu$ covers the target space $n$ times. 
Thus, 
the sum of the Euler characteristic 
of the set of the sheets on $\cM$ is  
\begin{equation}
 2-2g = n(2-2G)-i-2(t+h)-\sum_{k=1}^{|2-2G|}\omega_k,
  \label{Euler}
\end{equation}
since each of a branch point, a tube and a contracted handle 
contributes $-1$, $-2$ and $-2$ to the Euler characteristic, 
respectively. 
Here, $\omega_k$ is the contribution to the Euler characteristic 
from the $k$'th multiple branch point. 

\item (2) Numerical factor 

From the assumption that the position of the single branch points, 
the tubes and contracted handles can move on the target space, the factor  
$(\lambda A)^{i+t+h}/i!\,t!\,h!$ appears by integrating over 
their positions.%
\footnote{
$\lambda A$ is the only dimensionless expression of 
the area of the target space in this theory. 
} 
In addition, a tube and a contracted handle can attach to 
the sheets in $n(n-1)/2$ and $n/2$ possible ways, respectively. 
As a result, the numerical factor accompanied by $\nu$ is 
\begin{equation}
 \frac{\left(\lambda A\right)^{i+t+h}}{i!\,t!\,h!}
  \left(\frac{n(n-1)}{2}\right)^t\left(\frac{n}{2}\right)^h. 
  \label{factor}
\end{equation}

\item (3) Symmetry factor

Let us consider the fundamental group on 
$\cG\equiv\cM\backslash\{w_1,\cdots,w_i,z_1,\cdots,z_{|2-2G|}\}$. 
This group is generated by 
a set of loops $\{a_1,b_1,\cdots,a_G,b_G\}$ 
around the alpha and beta cycles of $\cM$, 
a set of loops $\{c_1,\cdots,c_i\}$ around the branch points, and 
a set of loops $\{d_1,\cdots,d_{|2-2G|}\}$ around the twists points,  
which are characterized by the relationship, 
\begin{equation}
 c_1\cdots c_i d_1\cdots d_{|2-2G|} 
  \prod_{j=1}^{G}a_jb_ja_j^{-1}b_j^{-1}=1. 
\end{equation}
We can define a homomorphism from the fundamental group 
on $\cG$ to the symmetry group $S_n$ by mapping a loop to 
a permutation of $n$-sheets.   
Thus, we see that the expression, 
\begin{equation}
 \sum_{s_1,t_1,\cdots,s_G,t_G \in S_n}\sum_{p_1,\cdots,p_i\in T_2}
  \frac{1}{n!}\delta(p_1\cdots p_i \Omega_n^{2-2G} 
  \prod_{j=1}^G s_j t_j s_j^{-1} t_j^{-1} ) 
  \label{symmetry}
\end{equation}
is equal to $\sum_{\nu \in \Sigma(G,n,i)}{1}/{|S_\nu|}$, 
where $|S_\nu|$ is the symmetry factor of the n-fold cover $\nu$. 
Note that the factor $n!$ in \eq{symmetry} is the number of 
possible labelings of $n$ sheets. 
\end{description}

Combining the above observations (1), (2) and (3), 
we see that the partition function \eq{chiral} is regarded as  
a sum over the covering maps with the properties described 
at the beginning of this section. 
This provides strong evidence 
for the equivalence of the large $N$ YM$_2$ 
and a two-dimensional string theory.



\end{document}